\documentclass[fontsize=6pt,journal=apchd5,manuscript=article]{achemso}
\usepackage{graphicx} 
\usepackage[utf8]{inputenc}
\usepackage[T1]{fontenc}
\usepackage{amsmath}
\usepackage{amsfonts}
\usepackage{amssymb}
\usepackage[dvipsnames]{xcolor}
\usepackage[numbers,sort&compress,super]{natbib}
\usepackage{multirow}
\usepackage{braket}
\usepackage{hyperref}
\usepackage{tabularx}
\usepackage{array}
\usepackage{graphicx}
\usepackage{colortbl}

\usepackage{bm}
\usepackage{enumitem}% http://ctan.org/pkg/enumitem

\usepackage{xspace} % bisogna stare attenti, crea casini. In alternativa wfq si usa come \wfq{} e via
\usepackage{tikz}
\usetikzlibrary{arrows, arrows.meta}
\usepackage{comment}
\usepackage{newtxtext}
\usepackage{mathtools}
\setkeys{acs}{articletitle = true}

\usepackage{mathptmx}

\newcommand*{\wfmu}{$\omega$F$\mu$\xspace}

\newcommand*{\wfqfmu}{$\omega$FQF$\mu$\xspace}

\newcommand*{\imm}{\mathop{}\!\mathrm{i}}

\newcommand*{\tqq}{\mathbf{T}^\mathrm{qq}}
\newcommand*{\tqmu}{\mathbf{T}^\mathrm{q\mu}}
\newcommand*{\tmuq}{\mathbf{T}^\mathrm{\mu q}}
\newcommand*{\tmumu}{\mathbf{T}^\mathrm{\mu\mu}}
\newcommand*{\tqfqqfq}{\mathbf{T}^\mathrm{\tilde{q}\tilde{q}}}

\newcommand*{\tmuqfq}{\mathbf{T}^\mathrm{\mu\tilde{q}}}

\newcommand*{\tqqfq}{\mathbf{T}^\mathrm{q\tilde{q}}}

\newcommand*{\agcluster}{Ag$_{164}$}
\newcommand*{\fq}{\tilde{q}}
\newcommand*{\tqfqqt}{\mathbf{T}^{\mathrm{\tilde{q}q},t}}
\newcommand*{\tqfqmut}{\mathbf{T}^{\mathrm{\tilde{q}\mu},t}}

\newcommand{\nextsubcolumn}[1][]{%
  \cr\noalign{\hfill}
  \if\relax\detokenize{#1}\relax\else\hsize=#1\setlength{\subcolumnwidth}{\hsize}\fi
}

\newcommand*{\sm}{SI}

\newlength{\subcolumnwidth}

\usepackage{xr-hyper}
\title[]{Atomistic Multiscale Modeling of Colloidal Plasmonic Nanoparticles}

\author{Luca Nicoli}
\affiliation{Scuola Normale Superiore,
             Piazza dei Cavalieri 7, 56126 Pisa, Italy.}
\author{Sveva Sodomaco}
\affiliation{Scuola Normale Superiore,
             Piazza dei Cavalieri 7, 56126 Pisa, Italy.} 
\author{Piero Lafiosca}
\affiliation{Scuola Normale Superiore,
             Piazza dei Cavalieri 7, 56126 Pisa, Italy.}
\author{Tommaso Giovannini}
\affiliation{Department of Physics, University of Rome Tor Vergata, Via della Ricerca Scientifica 1, 00133, Rome, Italy}
\email{tommaso.giovannini@uniroma2.it}          
\author{Chiara Cappelli}
\email{chiara.cappelli@sns.it}
\affiliation{Scuola Normale Superiore,
             Piazza dei Cavalieri 7, 56126 Pisa, Italy.}
\begin{document}
\maketitle

\begin{abstract}
A novel fully atomistic multiscale classical approach to model the optical response of solvated real-size plasmonic nanoparticles (NPs) is presented. The model is based on the coupling of the Frequency Dependent Fluctuating Charges and Fluctuating Dipoles (\wfqfmu), specifically designed to describe plasmonic substrates, and the polarizable Fluctuating Charges (FQ) classical force field to model the solvating environment. The resulting \wfqfmu/FQ approach accounts for the interactions between the radiation and the NP, as well as with the surrounding solvent molecules, by incorporating mutual interactions between the plasmonic substrate and solvent. \wfqfmu/FQ is validated against reference TD-DFTB/FQ calculations, demonstrating remarkable accuracy, particularly in reproducing plasmon resonance frequency shifts for structures below the quantum-size limit. The flexibility and reliability of the approach are also demonstrated by simulating the optical response of homogeneous and bimetallic NPs dissolved in pure solvents and solvent mixtures. 
\end{abstract}

\begin{center}
%\textbf{Keywords:} \textit{Atomistic simulations, solvatochromic-shifts, Silver, Gold, colloidal dispersions.}
\end{center}

% \begin{figure}
%     \centering
%     \includegraphics[width = \textwidth]{immagini/WhatsApp Image 2024-02-13 at 15.22.18.jpeg}]
%     \caption{Caption}
%     \label{fig:enter-label}
% \end{figure}
% \clearpage
% \newpage

\newpage

\section{Introduction}
In the past decades, colloidal nanoparticles (NPs), i.e. NPs dissolved in solution, have gained significant interest due to their applications in many technological contexts, such as sensing,\cite{mayer2011localized} biomedical applications,\cite{mcnamara2017nanoparticles} optoelectronics,\cite{dutta2020recent} and energy conversion.\cite{hu2010design} By choosing different precursors, reducing agents, solvents, and capping agents, nanostructured materials can be synthesized with a fine control of shape and size.\cite{yang2016colloidal} %Such physical features regulate the NP interaction with the external radiation, which can yield the excitation of the localized surface plasmon (LSP) typical of metal NPs.\cite{noguez2007surface} 
Size, shape, chemical composition, and the solvent can indeed affect the plasmon resonance frequency (PRF), i.e. the maximum of the NP absorption spectrum.\cite{chen2009shape} Such a feature is the basis of a particular class of sensors, which exploit the shift of the PRF upon change of the local refractive index (RI) of the solvent in which the plasmonic NPs are dissolved. Such devices have been widely employed in biosensing,\cite{lee2013highly,kim2018heteroassembled} where maximizing the induced PRF shift as a function of RI is crucial to enhance sensitivity.\cite{mayer2011localized,chen2008shape} 

% The sensing capabilities of such devices depend upon the ratio between the amount of shift that the LSP undergoes for a fixed variation of the local refractive index, which is also called sensitivity, and the peak linewidth, defining the so-called Figure of Merit (FOM): larger the shift and the smaller the linewidth, the greater the sensing capability.\cite{mayer2011localized}

%Within this framework, theoretical models for nanometrically engineered NPs and the effect of the local refractive index on their optical properties can support the rationalization of the plasmon band shift induced in colloidal NPs, potentially aiding the optimal design of high-sensitivity colorimetric LSP sensors.

%nanometrically engineered NP and the effect of the local refractive index on their plasmon band shift could help the rational design of high-sensitivity colorimetric LSP sensors.

Rationalizing solvent effects on the PRF of colloidal NPs is particularly challenging from a theoretical point of view. In fact, the optical properties of colloidal NPs result from the interplay of complex phenomena originating under the action of the external electric field, such as the appearance of a localized surface plasmon (LSP) excitation and the polarization of the solvent electron cloud. Solvent effects on PRFs result from a delicate balance between NP-solvent electrostatic (and polarization) interactions, charge-transfer effects, and the possible alteration of the plasmon decaying channels.\cite{foerster2017chemical} 

In principle, a proper description of all possible NP-solvent effects would require an \textit{ab initio} treatment of the whole system. However, first principle approaches become rapidly unfeasible due to their unfavorable scaling as a function of the system's size, thus hindering the simulation of realistic systems.
%In addition, the large dimensions of realistic metal NPs (thousands of atoms) and the even larger number of atoms of the solvent moiety needed to properly capture the NP-solvent interactions, add the final layer of complexity to the physics of such systems, effectively rendering them extremely difficult to model.
For these reasons, commonly exploited theoretical approaches to simulate plasmonic colloidal NPs are rooted in classical physics,\cite{steinbruck2011sensoric,szanto2021numerical,hu2016synthesis,hsiao2015colorimetric,yamada2022computational} generally making use of the classical Mie theory\cite{steinbruck2011sensoric}, the Boundary Element Method (BEM),\cite{szanto2021numerical} the Discrete Dipole Approximation (DDA),\cite{hu2016synthesis} and the Finite Difference Time Domain (FDTD).\cite{hsiao2015colorimetric} However, all these approaches are based on approximated descriptions of both the NP and the solvent, which is generally modeled as a continuum dielectric characterized by its specific permittivity $\varepsilon$. In addition, such approaches do not retain the atomistic nature of the system. %NPs and are thus not able to properly model how size, shape, chemical composition, and solvent effects can affect the PRF shift of generic colloidal NPs.\cite{chen2018morphology} 

%a fully atomistic modeling of such systems is still lacking.\\
In this paper, we propose a novel multiscale method where both the plasmonic NP and the solvent are treated at full atomistic level. In particular, we employ the fully atomistic electromagnetic model called Frequency Dependent Fluctuating Charges Fluctuating Dipoles (\wfqfmu)\cite{giovannini2022we} to describe the optical response of plasmonic NPs. Such a model is remarkably versatile and can be applied to NPs of any shape\cite{giovannini2022we} and chemical composition,\cite{nicoli2023fully} even at the quantum size limit ($<$ 5 nm).\cite{giovannini2022we} The solvent is modeled employing the polarizable Fluctuating Charges (FQ) force field\cite{rick1994dynamical,rick1995fluctuating,rick1996dynamical}, which is specifically designed to model the polarization of the electron cloud of molecular systems. For this reason, it has been widely exploited in the context of computational spectroscopy of solvated systems.\cite{giovannini2020molecular,giovannini2020theory,gomez2023multiple,giovannini2023continuum} 

The two approaches, \wfqfmu and FQ, are coupled in a multiscale fashion so that the resulting \wfqfmu/FQ model accounts for the mutual electrostatic interaction between the solvent and the plasmonic NP, allowing for the modeling of the optical properties of generic colloidal plasmonic NPs. \wfqfmu/FQ is also coupled to classical molecular dynamics (MD) simulations, which are exploited to sample the NP-solvent phase space. Therefore, differently from previous methods,\cite{steinbruck2011sensoric,steinbruck2011sensoric,szanto2021numerical,hu2016synthesis,hsiao2015colorimetric} the dynamical aspects of the solvation phenomenon, which are crucial to properly model solvent effects on spectroscopy, are taken into account.\cite{mennucci2019multiscale,gomez2023multiple,giovannini2020molecular} 

Note that, if solvent molecules are not adsorbed on the surface of the NP, the main solvent effect on plasmonic properties is the local RI variation of the medium surrounding the nanomaterial. This leads to a modification of the local optical field, generally causing a PRF redshift.\cite{xia2005shape,foerster2017chemical} Such effect is harnessed in many colorimetric-sensors for the detection of specific biomolecular analytes.\cite{lee2013highly,kim2018heteroassembled,loiseau2019core,piliarik2012high,mock2003local,underwood1994effect,steinbruck2011sensoric,rycenga2011controlling,haes2004unified,anker2008biosensing,stewart2008nanostructured}
In this work, we model the physical processes that lead to plasmon shift upon change of the local RI of the embedding medium. %To this end, we introduce a novel, fully atomistic approach that couples the \wfqfmu description of the nanostructure response with the FQ treatment of the solvent polarization. In the resulting \wfqfmu/FQ, mutual polarization effects between the solvent and the metal NP are introduced (see Fig. \ref{fig:schema-modello}). The plasmonic charges and dipoles of the \wfqfmu model generate a field that instantaneously polarizes the solvent FQ charges and vice versa. %Such molecular charges now act as an additional external electric field source to the plasmonic NP, effectively modifying its optical response. 

The paper is organized as follows: first, the novel \wfqfmu/FQ approach is presented after the theoretical foundations of \wfqfmu and FQ methods are recalled. Then, the computational protocol is presented and \wfqfmu/FQ is validated in comparison with reference Time-Dependent Density Functional Tight Binding/Fluctuating Charges (TD-DFTB/FQ\cite{lafiosca2022absorption}) results. The versatility and robustness of the approach are demonstrated by studying real-size homogeneous and bimetallic NPs dissolved in pure solvent or solvent mixtures. A summary and the main conclusions of the work end the manuscript.

\section{Theory}

\begin{figure} [!htbp]
    \centering
    \includegraphics[width = \textwidth]{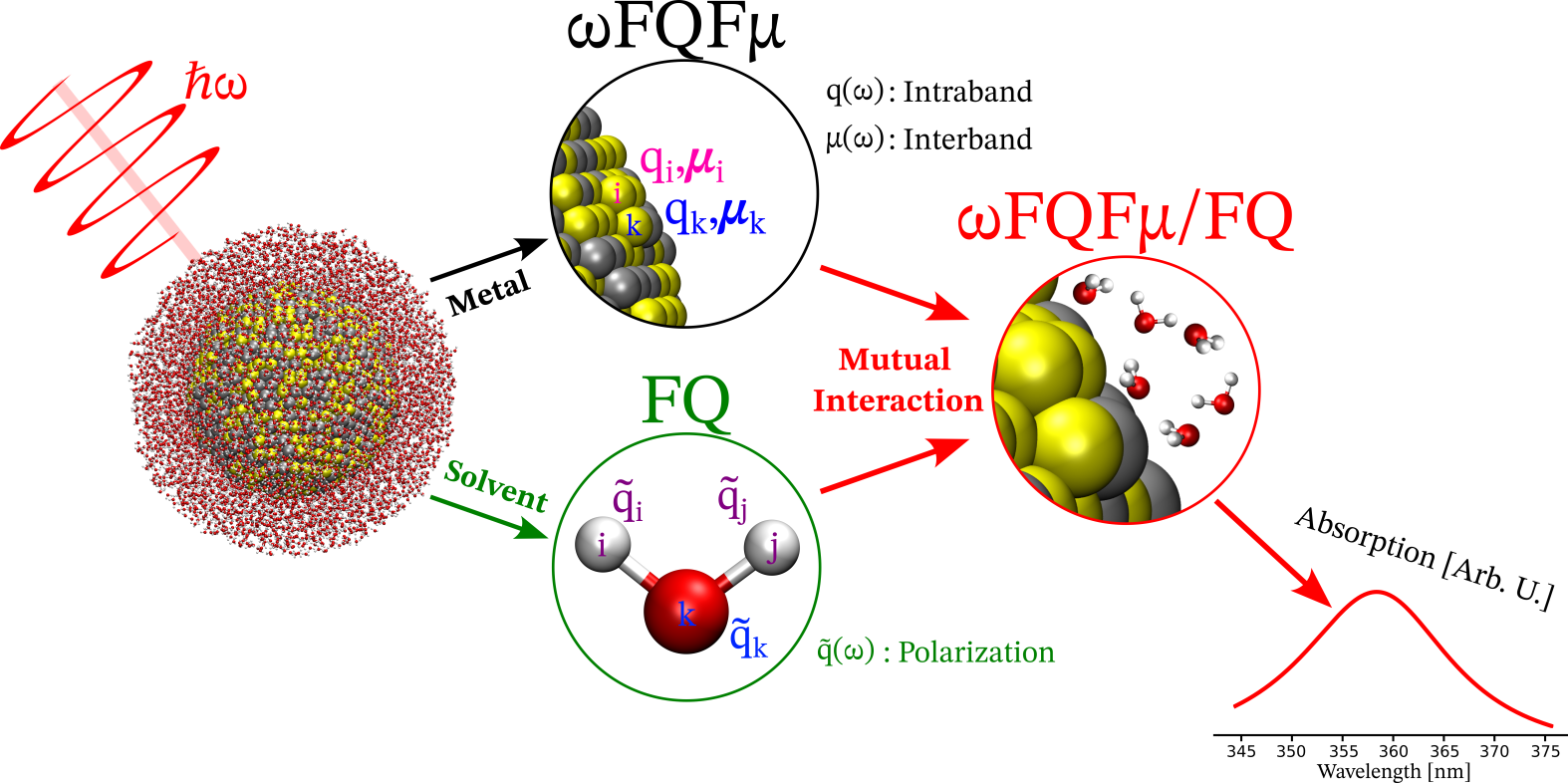}
    \caption{Pictorial view of the multiscale scheme employed to develop \wfqfmu/FQ.}
    \label{fig:schema-modello}
\end{figure}

In this section, \wfqfmu and FQ are briefly recalled and the \wfqfmu/FQ approach for modeling the optical properties of colloidal plasmonic NPs is presented (see Fig. \ref{fig:schema-modello}). 

\subsection{Plasmonic nanoparticle: the atomistic-electromagnetic \wfqfmu model}\label{sec:wfqfmu}

%CHIARA RECUPERA LE REFS QUI SOTTO!!!!!
%When an oscillating field interacts with an isolated metal NP, it can excite the LSP.\cite{maier2007plasmonics,besteiro2021theory} %which decay through intraband and interband mechanisms.
%To accurately model the optical response of plasmonic NPs, it is crucial to describe the physical mechanisms underlying the plasmonic excitations and the factors affecting them, such as shape, size, chemical compositions, and atomistic features.\cite{kelly2003optical,baumberg2022picocavities} To this end, we model the response of a plasmonic nanoparticle to an external monochromatic electric field through the \wfqfmu frequency-dependent atomistic electromagnetic model \cite{giovannini2019classical,giovannini2020graphene,giovannini2022we}. 
\wfqfmu models the plasmonic NP atomistically. Each atom is endowed with a set of complex frequency-dependent atom-centered charges $\mathbf{q}(\omega)$ ($\omega$FQs), and dipoles $\bm{\mu}(\omega)$ ({\wfmu}s) (see Fig. \ref{fig:schema-modello}), accounting for intraband and interband mechanisms, respectively. Charges are obtained by solving the following equation of motion, obtained by modulating a Drude-like conduction mechanics with quantum tunneling:\cite{giovannini2019classical,giovannini2020graphene}
\begin{equation}\label{eq:continuity-tunnelling}
-\imm\omega q_i(\omega) = \sum_j \left(\frac{A_jn_j(1-f_{ji}(l_{ij}))}{1/\tau_j-\imm\omega} + \frac{A_in_i(1-f_{ij}(l_{ij}))}{1/\tau_i-\imm\omega}\right)\frac{\phi_i(\omega)-\phi_j(\omega)}{l_{ij}}
\end{equation}
In eq.\ref{eq:continuity-tunnelling} $A_i$, $n_i$, and $\tau_i$ are the atomic effective area, the electron density, and the relaxation time associated with the intraband scattering events of the $i$-th atom, respectively. Quantum tunneling effects are expressed in terms of a Fermi-like damping function $f_{ij}(l_{ij})$, which exponentially damps the charge exchange between the atoms ($l_{ij}$ is the distance between $i$-th and $j$-th atoms).  %Sec. \ref{sec:si_wfqfmu} in the {\sm} for further details). 
$\phi_i(\omega)$ is the chemical potential of $i$-th atom, which reads:

\begin{equation}
   \phi_i(\omega) = V^{q}_i(\omega) + V^{\mu}_i(\omega) + V^\mathrm{ext}_i(\omega) \label{eq:potential-explicit}
\end{equation}
   
where $V^\mathrm{ext}_i$ is the electric potential associated with the optical radiation, whereas $V^{q}_i$ and $V^{\mu}_i$ are the electric potentials induced by charges and dipoles on the $i$-th atomic site (see Ref.\citenum{giovannini2022we} for more details).

%(see Sec. S1 in the \sm).
%
%\begin{align}
%    V^{q}_i(\omega)   & = \sum_k \tqq_{ik}q_k(\omega) \\
%    V^{\mu}_i(\omega) & = \sum_k \tqmu_{ik}\bm{\mu}_k(\omega)
%\end{align}
% \begin{align}
%     \phi_i(\omega) &=  \sum_k T^{\mathrm{qq}}_{ik} q_k(\omega) + \sum_k \tqmu_{ik}\bm{\mu}_k(\omega) + V^\mathrm{ext}_i(\omega) \label{eq:potential-explicit}\\
%     \mathbf{E}^{tot}_i(\omega) &= \sum_k\tmuq_{ik}q_k(\omega) + \sum_k\tmumu_{ik}\bm{\mu}_k(\omega) + \mathbf{E}^\mathrm{ext}_i(\omega) \label{eq:field-explicit}
% \end{align}
%
%
% It is important to note that to accurately depict the optical response of plasmonic subnanometer junctions and hot-spots, the incorporation of a phenomenological quantum tunneling description is essential \citep{scholl2013observation,giovannini2019classical,giovannini2022we,bonatti2022silico,esteban2012bridging,esteban2015classical}.
The plasmonic properties of alkali metals and graphene in the Pauli-blocking regime are correctly described by $\omega$FQs %($V^{\mu}_i(\omega) = 0$ in Eq. \ref{eq:potential-explicit}), 
which properly models intraband mechanisms.\cite{giovannini2019classical,giovannini2020graphene,bonatti2022silico}
However, when the interband absorption energy threshold is comparable to the plasmon resonance frequency (PRF), such as in noble metal nanoparticles, interband transitions become relevant to the decaying mechanism.\cite{maier2007plasmonics}
To model this process, an additional complex frequency-dependent dipole $\bm{\mu}(\omega)$ is assigned to each atom:
\begin{align}
\bm{\mu}_i(\omega) & = \alpha_i^{IB}(\omega)\ \mathbf{E}^\mathrm{tot}_i(\omega)\label{eq:dipoles} \\
                   & = \alpha_i^{IB}(\omega)\left[ \mathbf{E}^{q}_i(\omega) + \mathbf{E}^{\mu}_i(\omega) + \mathbf{E}^\mathrm{ext}_i(\omega) \right] \label{eq:field-explicit}
\end{align}
where $\alpha_i^{IB}(\omega)$ is the interband frequency-dependent polarizability of the $i$-th atom, and $\mathbf{E}^\mathrm{tot}_i(\omega)$ is the total electric field acting on the $i$-th dipole, accounting for charge-dipole ($\mathbf{E}^{q}$), dipole-dipole ($\mathbf{E}^{\mu}$), and dipole-field ($\mathbf{E}^{\text{ext}}$) interactions (see Ref. \citenum{giovannini2022we} for more details). %(see Sec. S1 in the \sm).
%
%\begin{align}
%    \mathbf{E}^{q}_i(\omega) & = \sum_k \tmuq_{ik}q_k(\omega) \\
%    \mathbf{E}^{\mu}_i(\omega) & = \sum_k \tmumu_{ik}\bm{\mu}_k(\omega)
%\end{align}
% \begin{align}
%     \phi_i(\omega) &=  \sum_k T^{\mathrm{qq}}_{ik} q_k(\omega) + \sum_k \tqmu_{ik}\bm{\mu}_k(\omega) + V^\mathrm{ext}_i(\omega) \label{eq:potential-explicit}\\
%     \mathbf{E}^{tot}_i(\omega) &= \sum_k\tmuq_{ik}q_k(\omega) + \sum_k\tmumu_{ik}\bm{\mu}_k(\omega) + \mathbf{E}^\mathrm{ext}_i(\omega) \label{eq:field-explicit}
% \end{align}
%
%
For homogeneous materials, $\alpha_i^{IB}(\omega)$ is determined from the frequency-dependent bulk-permittivity, whereas in the case of multimetallic systems, the interband polarizability of the $i$-th atom is expressed as a function of the local composition of the system. \cite{giovannini2022we,nicoli2023fully} 
The \wfqfmu charge-dipole coupled equations can be recast in the following set of complex linear equations:% (See Sec. \ref{sec:si_wfqfmu} for more details):
\begin{equation}\label{eq:total-system-general}
\begin{pmatrix} 
\mathbf{A}(\omega)\mathbf{T}^{\text{qq}} - \mathbf{Z}(\omega) & \mathbf{A}(\omega)\mathbf{T}^{\text{q}\mu}\\ \tmuq & \tmumu - \mathbf{Z}^\mathrm{IB}(\omega)  
\end{pmatrix}
\begin{pmatrix}
\mathbf{q} \\ \bm{\mu}
\end{pmatrix}
= 
\begin{pmatrix}-\mathbf{A}(\omega)\mathbf{V}^\mathrm{ext} \\ -\mathbf{E}^\mathrm{ext}
\end{pmatrix}
\end{equation}
where $\mathbf{A}(\omega)$ is a frequency-dependent matrix containing NP chemical and geometrical parameters, while $\mathbf{Z}(\omega)$, and $\mathbf{Z}^\mathrm{IB}(\omega)$ are diagonal matrices. %(See Sec. \ref{sec:si_wfqfmu} for their definitions). 
$\tqq$, $\tqmu$, $\tmuq$ and $\tmumu$ are charge-charge, charge-dipole, dipole-charge, and dipole-dipole interaction kernels, respectively (see Ref. \citenum{giovannini2019polarizable} for more details). 

When the linear system in Eq. \ref{eq:total-system-general} is solved, charges and dipoles modeling the optical intra- and interband response of the plasmonic NP are obtained. From such variables, the NP complex polarizability and the absorption cross-section $\sigma^{\mathrm{abs}}(\omega)$ can be computed (See Ref. \citenum{giovannini2022we} and Ref. \citenum{nicoli2023fully} for further details about the \wfqfmu model). %(See Sec. \ref{sec:si_wfqfmu} in the \sm). 
%\wfqfmu provides cost-effective modeling of the plasmonic response of a generic plasmonic NP, accurately reproducing fully quantum-mechanical results for isolated noble metal NPs, even in the quantum region size.\cite{giovannini2022we} Also, \wfqfmu atomistic nature makes it applicable to studying the optical response of plasmonic nanostructures with generic shapes and chemical composition, pending a reliable parametrization.\cite{nicoli2023fully,giovannini2022we}.

\subsection{Solvent: the Fluctuating Charges model (FQ)}\label{sec:fq}

The physics governing the optical response of solvent systems is utterly different from that of plasmonic materials. When external electric fields are applied, solvent molecules might experience several phenomena ranging from electronic transitions to molecular rovibrations depending on the external field frequency. In the following, we focus only on solvents that are transparent in the spectral region where the plasmonic nanostructure absorbs light. Note that this is generally the case for noble metal nanoparticles, whose PRF falls within the visible range (400 - 700 nm).\cite{litjens1999visible,sani2016spectral} In this way, the external fields exciting the plasmons have energies lower than those of the electronic transitions of the solvent molecules but sufficiently high to quench rovibrational effects completely. 
%In the modeling of the optical response of colloidal plasmonic NPs, we accurately choose solvents that are transparent in the visible range, such that 
% This is in general valid since we consider optical fields in the visible range (thus capable of exciting the LSP of noble metals) and solvents that are transparent in this window of frequencies
Thus, the interaction with the external field only yields the polarization of the solvent's electron cloud. This is modeled by using the Fluctuating Charges (FQ) force field,\cite{rick1994dynamical,rick1995fluctuating,rick1996dynamical} which has been extensively used in computational quantum chemistry for the modeling molecules in solution.\cite{giovannini2020molecular,nicoli2022assessing,gomez2023multiple} 
 Within FQ, each solvent atom is endowed with a charge $\tilde{\mathbf{q}}$ whose value is not fixed but can vary as a response to the external electric potential (see Fig. \ref{fig:schema-modello}). Such charge fluctuation is governed by the electronegativity equalization principle (EEP), \cite{mortier1985electronegativity,sanderson1951interpretation} which states that at equilibrium, each atom has the same electronegativity, i.e. the negative of the chemical potential, as reported by Parr.\cite{parr1983density} 
%Considering atomic partial charges as dynamical variables is well established since Parr has shown that the atomic electrons considered as an electron gas within Kohn-Sham theory possess .
%In a many-atom system, the full electron gas distributes equalizing the chemical potential, and this principle is called the 
%The FQ model is based on these concepts. 
The FQ energy, i.e. the energy required to create a partial charge on each atom is generally written as a second-order Taylor expansion in the charges. For a polyatomic system, this can be written as:\cite{rick1994dynamical}
\begin{equation}\label{eq:energy-fq-to-be-minimized}
    U(\tilde{\mathbf{q}}) = \sum_{\alpha,i} \left(\chi_{\alpha i}\tilde{q}_{\alpha i} + \frac{1}{2}\eta_{\alpha i}\fq^2_{\alpha i} + \tilde{V}^{tot}_{\alpha i}\fq_{\alpha i}\right)+ \sum_{\beta,k < \alpha,i}\tilde{q}_{\beta k}T^{\tilde{q}\tilde{q}}_{\beta k, \alpha i}\tilde{q}_{\alpha i}
\end{equation}
where Greek and Roman indices run over molecules and atoms, respectively. $\chi_{\alpha i}$ is the electronegativity of the $i$-th atom of the $\alpha$-th molecule, and  $\mathbf{T}^{\fq\fq}$ is the charge-charge interaction kernel\cite{giovannini2019polarizable}  %(See Eq. \ref{eq:si_fq_kernel} in the \sm) 
of which the diagonal elements $\eta_{\alpha i}$ are chemical hardnesses, representing self-interaction polarization term. Both electronegativity and chemical hardness are well rooted in Conceptual Density Functional Theory \cite{geerlings2003conceptual} and are the only free parameters defining the model. Moreover, $\tilde{\mathbf{V}}^{tot}$ is the total external electric potential on each atomic site, which in the FQ case is the potential associated with the external optical field ($\tilde{\mathbf{V}}^{tot} = \tilde{\mathbf{V}}^{ext}$). %The electronegativity per unit charge for each atom is given as
%\begin{equation}
%    \tilde{\chi}_{\alpha i} = \left(\frac{\partial U}{\partial \tilde{q}_{\alpha i}}\right)
%\end{equation}
%The FQ charges are those for which all electronegativities are equal, thus satisfying the EEP.
To constrain the charge of each molecule $Q^{tot}_{\alpha}$ to a constant, a set of Lagrange multipliers $\lambda_{\alpha}$ is exploited. The energy functional in Eq. \ref{eq:energy-fq-to-be-minimized} thus becomes:
\begin{equation}\label{eq:total-fq-functional}
     F(\tilde{\mathbf{q}}, \bm{\lambda}) = U(\tilde{\mathbf{q}}) + \sum_{\alpha}\lambda_{\alpha}\left(\sum_i (\fq_{\alpha i}) - Q^{tot}_{\alpha}\right)
\end{equation}

From the minimization of Eq. \ref{eq:total-fq-functional} with respect to the variables ($\tilde{\mathbf{q}}$ and $\bm{\lambda}$), the FQ polarization equations are obtained, implying that the electronic degrees of freedom of the solvent instantaneously rearrange without energy dissipation.\cite{rick1994dynamical}
%
%obtains a set of coupled linear equations:
%
%\begin{align} 
%& \sum_{\beta k} T^{\fq\fq}_{\alpha i, \beta k}\fq_{\beta k}  + \lambda_{\alpha} = - \chi_{\alpha i} - \tilde{V}^{tot}_{\alpha i}\label{eq:fq-master}\\
%&\sum_{i} \fq_{\alpha i} = Q^{tot}_{\alpha}\label{eq:fq-lagrangian-multipliers-eom}
%\end{align}
%These are generally known as the FQ polarization equations and they imply that the electronic degrees of freedom of the solvent always rearrange instantaneously without energy dissipation.\cite{rick1994dynamical} 
%For the sake of completeness, Eq. \ref{eq:fq-master} and Eq. \ref{eq:fq-lagrangian-multipliers-eom} can be seen from another physical perspective to emerge from the equation of motion of fictitious partial atomic-charges (from a Lagrangian formalism) under the assumption that the electronic degrees of freedom of the solvents always rearrange instantaneously with the constraint of fixed molecular charges.\cite{rick1994dynamical} 
%
Such an approximation is also valid for optical fields in the visible range and for transparent solvents.\cite{cappelli2016integrated,giovannini2020molecular,giovannini2020theory} 
When the total electric field is monochromatic at frequency $\omega$, the FQ polarization equations in the frequency domain read as follows (See Sec. S1.1 in the Supporting Information \sm \  for further details):  
%
%it is useful to cast the FQ polarization equations in the frequency domain:
%\begin{align}
%    &\sum_{\beta k} T^{\fq\fq}_{\alpha i, \beta k}\fq_{\beta k}(\omega)  + \lambda_{\alpha}(\omega) = -\tilde{V}^{tot}_{\alpha i}(\omega) \label{eq:fq-polarization-equation-freq}\\
%    &\sum_{i} \fq_{\alpha i}(\omega) = 0 \label{eq:fq-lagrangian-multipliers-freq}
%\end{align}

%Equations \ref{eq:fq-polarization-equation-freq} and \ref{eq:fq-lagrangian-multipliers-freq} can be written in a useful matrix notation:
%
\begin{equation}
    \begin{pmatrix}
        \mathbf{T}^{\fq\fq} & \mathbf{1}\\\mathbf
        {1}^t & \mathbf{0}
    \end{pmatrix}
    \begin{pmatrix}
    \tilde{\mathbf{q}}(\omega) \\ \bm{\lambda}(\omega)    
    \end{pmatrix} = 
    \begin{pmatrix}
         -\tilde{\mathbf{V}}^{ext}(\omega) \\
        0
    \end{pmatrix}
    \label{eq:fq_response_w}
\end{equation}
where $\mathbf{1}$ is a rectangular matrix containing the blocks associated with the Lagrange multipliers. By solving Eq. \ref{eq:fq_response_w}, the atom-centered FQ charges, modeling the instantaneous polarization of the electronic cloud of the solvent under the application of an external monochromatic field, are obtained. %The value of such charges depends on the functional form chosen for the charge-charge interaction kernel ($\tqfqqfq$) and from the numerical values of the chemical hardness ($\eta$) of each atom species. The value of the electronegativity ($\chi$) only influences the static component of the solvent polarization charges.

\subsection{Optical response of a plasmonic nanoparticle in solution: the $\omega$FQF$\mu$/FQ model}

%As stated above, the optical response of plasmonic materials can significantly be affected by a surrounding solvent.\cite{foerster2017chemical,ghosh2004solvent,nath2010ligand} 
%CHIARA QUI!!!!! SPOSTARE????If solvent molecules are not adsorbed on the surface of the NP, the main solvent effect on plasmonic properties is the local refractive index (RI) variation of the medium surrounding the nanomaterial. This leads to a modification of the local optical field, generally causing a PRF redshift.\cite{xia2005shape,foerster2017chemical} Such effect is harnessed in many colorimetric-sensors for the detection of specific biomolecular analytes.\cite{lee2013highly,kim2018heteroassembled,loiseau2019core,piliarik2012high,mock2003local,underwood1994effect,steinbruck2011sensoric,rycenga2011controlling,haes2004unified,anker2008biosensing,stewart2008nanostructured}
%CHIARA QUI!!!!! QUESTA ROBA E' RIPETUTA!!!!! In this work, we only model the physical processes that lead to plasmon shift upon change of the local RI of the embedding medium. To this end, 

 \wfqfmu for describing the nanostructure's response and FQ for modeling solvent polarization are coupled in a multiscale fashion. In the resulting \wfqfmu/FQ approach, mutual polarization effects between the solvent and the metal NP are introduced (see Fig. \ref{fig:schema-modello}). 
 To this end, we include the solvent-induced electric potential $\mathbf{V}^{\fq}(\omega)$ and electric field $\mathbf{E}^{\fq}(\omega)$ acting on the NP's charges $\mathbf{q}(\omega)$ and dipoles $\bm{\mu}(\omega)$ respectively. Thus, the total chemical potential and field (Eq. \ref{eq:potential-explicit} and Eq. \ref{eq:field-explicit} respectively) acting on each atomic site now read:

\begin{align}
    \phi_i(\omega) &= V^{q}_i(\omega) + V^{\mu}_i(\omega) 
 +  V^{\fq}_i(\omega)+ V^\mathrm{ext}_i(\omega) \label{eq:potential-explicit-si}\\
    \mathbf{E}^{tot}_i(\omega) &= \mathbf{E}^{q}_i(\omega) + \mathbf{E}^{\mu}_i(\omega) + 
 \mathbf{E}^{\fq}_i(\omega) +\mathbf{E}^\mathrm{ext}_i(\omega) \label{eq:field-explicit-si}
\end{align}

The electric potential and field induced by the FQ solvent charges $\tilde{\mathbf{q}}(\omega)$ on the NP's atomic sites read:
\begin{align}
    V^{\fq}_i(\omega) &= \sum_{\alpha,k}T^{q\fq}_{i, \alpha k}\fq_{\alpha k} (\omega) \label{eq:si_pot_fq}\\ 
    \mathbf{E}^{\fq}_i(\omega) &= \sum_{\alpha,k}\mathbf{T}^{\mu\fq}_{i,\alpha k}\fq_{\alpha k}(\omega)\label{eq:si_field_fq}
\end{align}

$\mathbf{T}^{q\fq}$ and $\mathbf{T}^{\mu\fq}$ are the Coulomb interaction kernels between the FQ solvent charges and \wfqfmu plasmonic charges and dipoles respectively, i.e.:
%The Coulomb interaction kernel elements between FQ solvent charges and the \wfqfmu plasmonic charges ($\mathbf{T}^{q\fq}$) and dipoles ($\mathbf{T}^{\mu\fq}$) read:
\begin{align}
    T_{ij}^{q\tilde{q}}&=\frac{1}{|\mathbf{r}_{ij}|} \label{eq:si_twfq_fq}\\ %&& \tqqfq = \tqfqq^{,T}\\
\bm{T}_{ij}^{\mu\tilde{q}}&=\mathbf{\nabla}_{r_i}T_{ij}^{q\tilde{q}} \label{eq:si_twfmu_fq}%&& \tqfqmu = \tmuqfq^{,T}
\end{align}
where $\mathbf{r}_{ij}$ is the distance between atoms $i$ and $j$.

To account for mutual polarization effects, the total potential $\tilde{\mathbf{V}}^{\mathrm{tot}}$ acting on the solvent atomic sites includes the potential generated by the NP's charges $\tilde{\mathbf{V}}^{q}(\omega)$ and dipoles $\tilde{\mathbf{V}}^{\mu}(\omega)$:

\begin{equation}\label{eq:pot_fq_wfqfmu}
    \tilde{V}^{tot}_{\alpha i}(\omega) = \tilde{V}^{q}_{\alpha i}(\omega) +\tilde{V}^{\mu}_{\alpha i}(\omega)
\end{equation}

where the electric potential and field induced by the \wfqfmu charges and dipoles on the $i$-th solvent atom of the $\alpha$-th molecule read:
\begin{align}
    \tilde{V}^{q}_{\alpha i}(\omega) &= \sum_k T^{\fq q}_{\alpha i, k}q_k(\omega) \label{eq:si_pot_wfq}\\
    \tilde{V}^{\mu}_{\alpha i}(\omega) &= \sum_k \mathbf{T}^{\fq \mu}_{\alpha i, k} \bm{\mu}_k(\omega)\label{eq:si_field_wfq}
\end{align}
Note that in Eq. \ref{eq:pot_fq_wfqfmu} we neglect local field effects (i.e. $\tilde{{V}}_{\alpha i}^{\mathrm{ext}}(\omega) = 0$). 
 % The plasmonic charges and dipoles of the \wfqfmu model generate a field that instantaneously polarizes the solvent FQ charges and vice versa. %Such molecular charges now act as an additional external electric field source to the plasmonic NP, effectively modifying its optical response. 
% The \wfqfmu equations are thus modified to take into account the potential % (see Eq. \ref{eq:potential-explicit}) 
% and field %(see Eq. \ref{eq:field-explicit}) 
% generated by the FQ charges ($\textbf{V}^{\fq}(\omega)$ and $\textbf{E}^{\fq}(\omega)$, respectively. CHIARA QUI!!!!, see also Eqs. \ref{eq:si_pot_fq} and \ref{eq:si_field_fq} in the \sm).
% Analogously, the total potential $\tilde{\mathbf{V}}^{\mathrm{tot}}$ acting on the solvent FQs is modified to account only for the potential induced by the NP charges $\tilde{\mathbf{V}}^{q}(\omega)$ and dipoles $\tilde{\mathbf{V}}^{\mu}(\omega)$ (see Eqs. \ref{eq:si_pot_wfq} and \ref{eq:si_field_wfq} in the \sm), thus neglecting local field effects (i.e. $\tilde{\mathbf{V}}^{\mathrm{ext}}(\omega) = 0$ in Eq. \ref{eq:fq_response_w}). 

The coupled \wfqfmu and FQ equations can thus be recast in a linear system defining the \wfqfmu/FQ master equations:
\begin{equation}\label{eq:total-system-final}
\begin{pmatrix} 
\mathbf{A}(\omega)\tqq - \mathbf{Z}(\omega) & \mathbf{A}(\omega)\tqmu & \mathbf{A}(\omega)\tqqfq & \mathbf{0}\\ \mathbf{T}^{q\mu} & \tmumu - \mathbf{Z}^\mathrm{IB}(\omega) & \tmuqfq  & \mathbf{0} \\
\tqfqqt & \tqfqmut & \tqfqqfq & \mathbf{1} \\
\mathbf{0} & \mathbf{0} & \mathbf{1}^t & \mathbf{0}
\end{pmatrix}
\begin{pmatrix}\mathbf{q} \\ \bm{\mu} \\ \tilde{\mathbf{q}}\\ \bm{\lambda}
\end{pmatrix}
= 
\begin{pmatrix} -\mathbf{A}(\omega)\mathbf{V}^\mathrm{ext}(\omega) \\ -\mathbf{E}^\mathrm{ext}(\omega) \\
0 \\ 0
\end{pmatrix}
\end{equation}

% In eq.\ref{eq:total-system-final} $\mathbf{T}^{q\fq}$, $\mathbf{T}^{\mu\fq}$, are the Coulomb interaction kernels between the solvent polarization charges and the plasmonic charges and dipoles respectively (see Eqs. \ref{eq:si_twfq_fq} and \ref{eq:si_twfmu_fq} in the \sm).
By solving Eq.\ref{eq:total-system-final}, the charges and dipoles defining the response of the nanostructure as modified by the presence of the solvent are obtained. This allows the simulation of the optical response of plasmonic substrates with arbitrary shape and chemical composition embedded in a generic solvent or solvent mixture  (See Sec. S1.2 in the \sm).

\section{Computational Protocol}\label{sec:comp_prot}

\begin{figure}[!htbp]
    \centering
    \includegraphics[width = \textwidth]{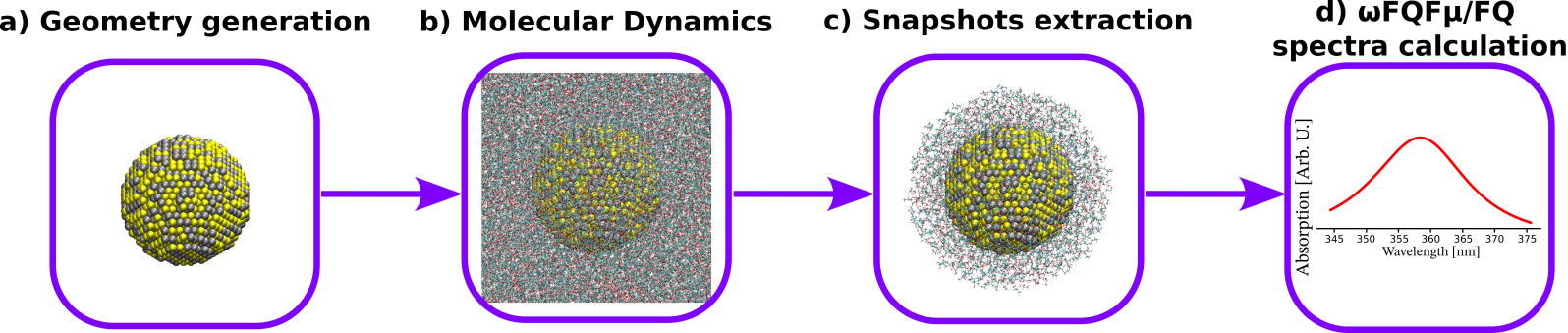}
    \caption{Graphical scheme of the computational protocol employed to compute \wfqfmu/FQ absorption spectra of colloidal plasmonic NPs (see also Secs. S2.1-S2.4 in the {\sm}).}
    \label{fig:protocol}
\end{figure}
%Solvation phenomena affect the absorption properties of plasmonic nanostructures \cite{ghosh2004solvent, nath2010ligand}, and as stated above, \wfqfmu/FQ is developed to address the LSP shift due to the change in the local refractive index of the medium embedding the plasmonic NP. 

% Such effect is caused by the mutual electrostatic interaction between the solvent molecules and the plasmonic NP. 
% Notably, the experimentally measured solvent-induced LSP shift is averaged over the multitude of configurations that the solvent molecules take around the nanoparticle during the time of measurement. 
In this work, we apply \wfqfmu/FQ to the calculation of the absorption cross-section of plasmonic nanoparticles in solution. To reproduce the experimental PRF shift induced by solvent effects, the dynamic nature of the solvation phenomena needs to be properly described. To this end, we adapt the protocol designed for molecular systems in solution\cite{giovannini2020molecular} to the specific case of colloidal plasmonic NPs. 
%
%Notably, the experimentally measured LSP shift is generally averaged over the multitude of configurations that the solvent molecules take around the nanoparticle during the time of measurement. Thus, to reproduce the experimental LSP shifts, we also need to adequately account for the dynamic nature of the solvation phenomena, i.e., the spectrum we compute needs to be averaged over a set of solvent-NP configurations. 

% To this end, we deploy a computational protocol specifically designed for this task (See Fig. \ref{} and Sec. S XXX), that can be divided in steps:
% \begin{enumerate}
%     \item \textbf{NP structure generation} -- First of all, by employing a in-house code, we generate the geometry of the isolated plasmonic NP (See Fig. \ref{}-a  and Sec. S XXX in the SI);
%     \item \textbf{Molecular Dynamics simulation} -- Then, we solvate the NP and we perform the  Molecular Dynamics (MD) simulations of the colloidal NP (See Fig. \ref{}-a  and Sec. S XXX in the SI );
%     \item \textbf{Snapshots extraction} -- From the MD we then extract uncorrelated representative structures of the whole system. For each structure, we then carve a spherical ''droplet'' of solvent molecules and metal atoms centered on the NP, which we call snapshot (See Fig. \ref{}-a  and Sec. S XXX in the SI) 
%     \item \textbf{\wfqfmu/FQ calculations and extraction of the absorption spectrum} -- For each snapshot we then perform \wfqfmu/FQ calculation. The overall spectroscopic property is then computed as average over the snapshots (See Fig. \ref{} and Sec. S XXX in the SI).
% \end{enumerate}
%
\noindent The protocol can be divided into four main steps (see Fig. \ref{fig:protocol}): 
\begin{enumerate}[label=(\roman*)]
    \item \textit{Geometry generation}: The geometry of isolated plasmonic NPs is generated by using an in-house code that employs the Atomic Simulation Environment (ASE) Python module v. 3.17 \cite{larsen2017atomic} (see Fig. \ref{fig:protocol}a and Sec. S2.1 in the \sm)
    \item \textit{Molecular Dynamics}: The nanostructure is solvated. To sample the NP-solvent phase-space, we perform a classical Molecular Dynamics (MD) simulation of the solvated colloidal system by using the GROMACS software package (version 2020.4) and a suitable force-field\cite{abraham2015gromacs} (see Fig. \ref{fig:protocol}b and Sec. S2.2 in the \sm).
    \item \textit{Extraction of structures}: From the MD trajectory, we extract 25 uncorrelated representative structures of the whole system. The number of structures ensures convergence of the spectral signal (see Sec. S2.4.2 in the \sm). For each structure, we retain all solvent molecules that are at most 15 \AA~ from the NP surface, resulting in a spherical droplet (see Fig. \ref{fig:protocol}c) and Sec. S2.4.1 in the \sm).
    \item \textit{\wfqfmu/FQ spectral calculations}: For each spherical droplet, absorption cross sections are computed at the \wfqfmu/FQ level. The overall spectroscopic response is then recovered as the average over all structures (see Fig. \ref{fig:protocol}d and Sec. S2.3 in the \sm). All \wfqfmu/FQ calculations are performed by employing a stand-alone Fortran 95 package.
\end{enumerate}

\section{Results and Discussion}

In this section, \wfqfmu/FQ is applied to compute the absorption properties of noble metal NPs in solution. First, the model is validated by reproducing reference data. Then, it is employed to study solvent effects on realistic homogeneous and bimetallic NPs dissolved in a pure solvent or a solvent mixture, showcasing the potential and flexibility of the approach.

\subsection{Model Validation}
%Optical properties of small \agcluster cluster in water: \wfqfmu/FQ model versus TD-DFTB/FQ \textit{ab-initio} model}

The \wfqfmu/FQ approach is validated against vacuo-to-water PRF shifts of a small silver spherical-like cluster composed of 164 atoms (\agcluster). \wfqfmu/FQ values are compared to polarizable Time-Dependent Density Functional Tight Binding/Fluctuating Charges (TD-DFTB/FQ) calculations.\cite{lafiosca2022absorption} %TD-DFTB/FQ is a hybrid \textit{ab initio}/classical approach which couples TD-DFTB with the polarizable FQ force field, thus accounting for mutual NP-solvent polarization. 
The advantageous computational scaling of TD-DFTB/FQ, which is reached through the approximation of two-electron interactions, makes it capable of handling larger systems than Time-Dependent Density Functional Theory-based methods (TD-DFT) while preserving accuracy.\cite{lafiosca2022absorption}
% Comparing \wfqfmu/FQ with TD-DFTB/FQ thus highlights accuracy of our fully atomistic model.  
The initial geometry of {\agcluster} is taken from Ref. \citenum{liu2020td}: from that structure, MD simulations in aqueous solution are run, a single random snapshot is extracted and cut in a spherical droplet containing water (WAT) molecules within 10 {\AA} from the surface of the NP, resulting in a total number of 807 water molecules. Such distance is chosen to account for the most relevant NP-water interactions (See Sec. S2.4.1 in the \sm).
The absorption spectrum of {\agcluster} in vacuo (i.e. by removing all water molecules from the snapshot) and in aqueous solution (\agcluster/WAT$_{807}$) are computed by using \wfqfmu and DFTB, and \wfqfmu/FQ and DFTB/FQ, respectively (see Sec. S2.5 in the {\sm} for more details on the DFTB/FQ calculations). 
% In particular, we used standard DFTB for the calculations of \agcluster in vacuo. For the calculations in the solution, we employed the DFTB/FQ model.
To evaluate quantum effects at the NP-solvent interface, we also performed DFTB/DFTB$_{WAT}$/FQ calculations, where the first solvation shell is treated at the DFTB level of theory (109 water molecules), whereas the remaining 698 WAT molecules are described at the FQ level. 
\wfqfmu parameters are recovered from Ref. \citenum{giovannini2022we}. The FQ WAT molecules are modeled by using the parameters reported in Ref. \citenum{rick1994dynamical} (WAT$_1$, see Tab. S1 in Sec. S2.3 in the \sm). Note that additional calculations were performed by using the FQ parameters reported in Refs. \citenum{carnimeo2015analytical,giovannini2019effective,ambrosetti2021quantum} (see Sec. S3 in the \sm). 

% \begin{figure}[!h]
%     \centering
%     \includegraphics[width = 0.45\textwidth]{immagini/immagine_wat_dftb-delta.png}
%     \caption{Absorption spectra of \agcluster in vacuo (VAC) and aqueous solution (WAT) as calculated at the DFTB/FQ (A), DFTB/DFTB$_{WAT}$/FQ (B), and \wfqfmu/FQ (C) using WAT$_1$ FQ parameterization (see right panel for a graphical representation of a sperical droplet).\cite{rick1994dynamical} The solvatochromic shifts ($\Delta E$) are also given. All spectra are normalized to the maximum in vacuo.} %$\Delta E$ reports the computed vacuo-to-water PRF shift. On the right panel, a graphical depiction of the structure used for each calculation is presented and the elements are colored according to the level of theory used (orange-DFTB, gray-\wfqfmu, blue-FQ). In the green inset a zoom of the DFTB/DFTB$_{WAT}$/FQ structure is reported to highlight the presence of water molecules treated at the DFTB level of theory (orange).}
%     \label{fig:wat_dftb_fq}
% \end{figure}

\begin{figure}[!t]
    \centering
    \includegraphics[width = 0.8\textwidth]{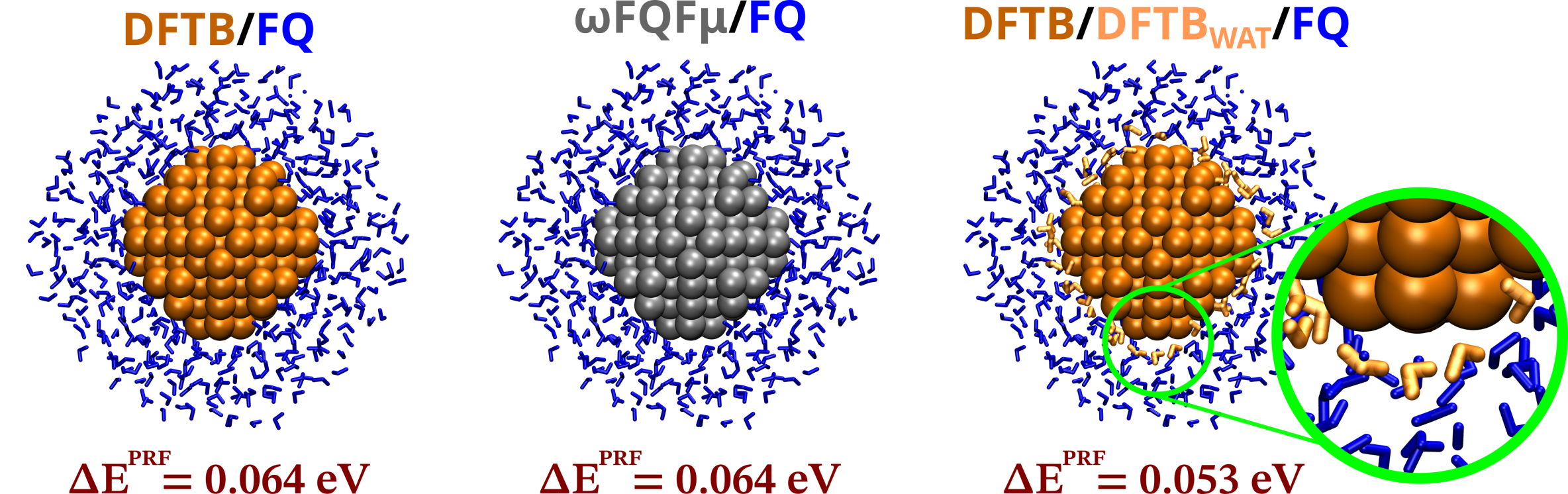}
    \caption{Vacuo-to-water PRF shifts (in eV) computed by using DFTB/FQ, \wfqfmu/FQ, and DFT/DFTB$_{WAT}$/FQ levels of theory. A graphical depiction of the structure used for each calculation is presented and the elements are colored according to the level of theory used (orange-DFTB, gray-\wfqfmu, blue-FQ). In the green inset a zoom of the DFTB/DFTB$_{WAT}$/FQ structure is reported to highlight the presence of water molecules treated at the DFTB level of theory (orange).}
    \label{fig:wat_dftb_fq}
\end{figure}

% Computed absorption spectra are graphically depicted in Fig. \ref{fig:wat_dftb_fq}, which shows spectra calculated at the DFTB/FQ level (panel A), DFTB/DFTB$_{WAT}$/FQ (panel B), and \wfqfmu/FQ (panel C). The corresponding PRF shifts ($\Delta E = E^{VAC} - E^{WAT}$) are also indicated (in eV).
In Fig. \ref{fig:wat_dftb_fq}, we report the PRF shifts ($\Delta E^{\text{PRF}} = E^{VAC} - E^{WAT}$) in eV calculated at the DFTB/FQ level (panel A), \wfqfmu/FQ (panel B), and DFTB/DFTB$_{WAT}$/FQ (panel C). 

All methods predict a very similar vacuo-to-water PRF redshift. The sign of the shift is expected due to the increase of the refractive index of the NP surrounding medium, as also supported by experimental observations.\cite{zhang2016noble,lee2011refractive} Notably, the hybrid DFTB/FQ and the fully classical \wfqfmu/FQ approaches predict the same solvatochromic shift (64 meV). This is particularly remarkable because the dimension of the studied NP falls within the quantum size region, where quantum effects are expected to play a major role. Our results highlight the accuracy of \wfqfmu in describing such structures.\cite{liebsch1993surface}
% PRF shift proves to be rather sensitive to the FQ parameterization e.g. for DFTB/FQ  the PRF shift ranges from 0.061 eV (FQ$_2$) to 0.143 eV (FQ$_4$).
When the first solvation shell of water molecules is described at the DFTB level (DFTB/DFTB$_{WAT}$/FQ), the predicted solvatochromic shift decreases by $\sim$ 10 meV (53 meV). Such reduction can be attributed to the inclusion of purely quantum NP-solvent interactions, mainly related to Pauli repulsion effects, \cite{giovannini2019quantum,amovilli2020effect,lafiosca2022absorption} which are not taken into account by both DFTB/FQ and \wfqfmu/FQ. 
Remarkably, the computational cost associated with \wfqfmu/FQ is negligible as compared to DFTB-based methods (see Tab. S4 in the \sm). Indeed, the favorable computational scaling of \wfqfmu/FQ provides a substantial 99.4 \% speed-up of the calculation with respect to DFTB/FQ, thus representing an effective, reliable, and cost-effective alternative to state-of-the-art \emph{ab initio} methods.

%To conclude, in this section, we validated \wfqfmu/FQ against the reproduction of the vacuo-to-water PRF shift of a small silver cluster (\agcluster). The agreement with DFTB/FQ results for such a small-sized system, which is in the limit of quantum confinement, is remarkable. In addition to this, the classical nature of \wfqfmu/FQ ensures good performances for larger NPs in the classical limit, while the favorable computational scaling of \wfqfmu/FQ provides a substantial 99.4 \% speed-up of the calculation with respect to DFTB/FQ. 

% Summing up, \wfqfmu/FQ is capable of reproducing \textit{ab-initio} TD-DFTB/FQ vacuo-to-water LSP shifts of a small, in the limit of quantum confinement, silver nanostructure (\agcluster), providing a substantial 99.4 \% speed-up of the calculations.

\subsection{Homogeneous colloidal NPs}\label{sec:risultati-2}

\begin{figure}[!htbp]
    \includegraphics[width=0.5\textwidth]{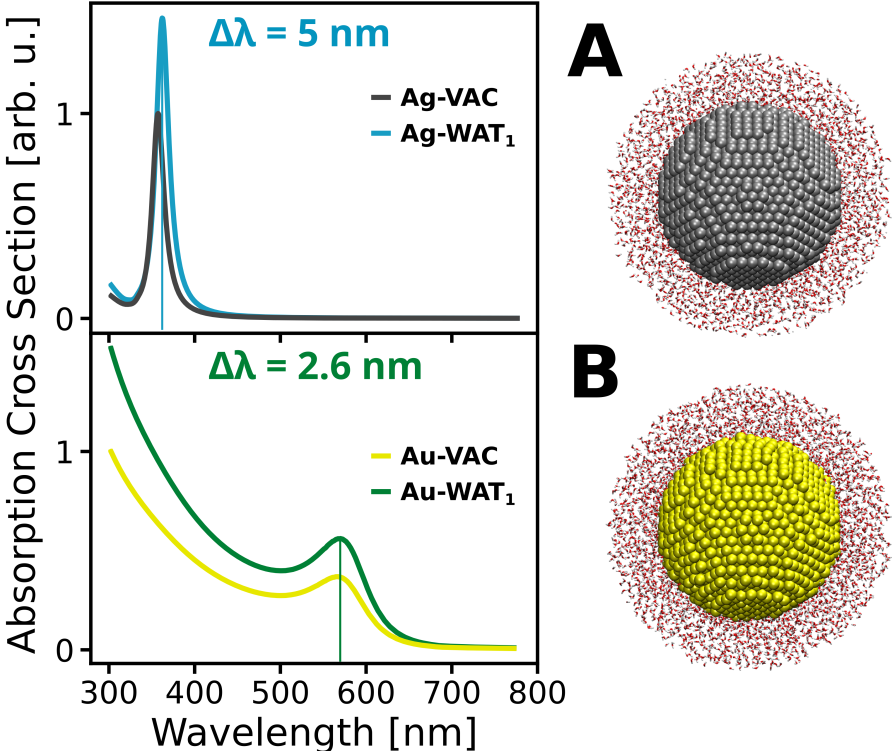}
    \caption{\wfqfmu/FQ absorption spectra of (A) Ag$_{3851}$ and (B) Au$_{3851}$ in vacuo (VAC) and water (WAT$_1$). $\Delta \lambda$ is the vacuo-to-water solvatochromic shift in nm. All spectra are normalized to the corresponding maximum in vacuo.} %In the right panel, we report a graphical depiction of one of the snapshots employed for the calculation of the spectrum, obtained by removing half of the water molecules for better visualization.}
    \label{fig:spheres}
\end{figure}

The favorable computational scaling of \wfqfmu/FQ opens up to computing the plasmonic response of large NPs in solution. We first consider homogeneous silver and gold spherical NPs (diameter = 5 nm, 3851 atoms) in aqueous solution, as modeled by using the WAT$_1$ parameterization\cite{rick1994dynamical}.
%
%leverage both the favorable computational scaling of \wfqfmu/FQ, as well as its atomistic nature, to compute the optical properties of large plasmonic NPs of complex chemical composition in solution. In particular, we first 
%
%for which experimental data are available. 
% In this section, we highlight how \wfqfmu/FQ is the most general model available in the literature for the study of the optical properties of plasmonic NPs in solution.
% We begin presenting how \wfqfmu/FQ is capable of computing the optical properties of homogeneous spherical NPs, highlighting the dimensions and the fact that we can keep the right $\Delta \lambda$ between silver and gold.
% Then we highlight how \wfqfmu/FQ is capable of computing the optical properties of much more complex geometries and that there is no other model capable of doing this.
The average number of WAT molecules in the FQ region is 6464 (19392 atoms). Computed absorption cross-sections both in vacuo (VAC) and in water (WAT) are plotted in Fig. \ref{fig:spheres}A-B, respectively, together with the corresponding vacuo-to-water PRF shifts ($\Delta \lambda$, in nm). 
The spectra of both systems are characterized by a main plasmonic band centered at about 360 nm (Ag NPs) and 570 nm (Au NPs). Au peaks are broader than Ag bands, in agreement with previous observations.\cite{hottin2013comparison,dengler2012near} In the studied region (300-800 nm), the spectra of Au systems are also characterized by the presence of a broad and intense band associated with interband absorption.\cite{giovannini2022we} 
When dissolved in solution, the main spectral features are maintained. However, we note that the plasmonic peak increases in intensity and slightly redshifts, in line with experimental data.\cite{chen2008shape,ghosh2004solvent, zhang2016noble, lee2011refractive} For Ag, \wfqfmu/FQ predicts a red-shift of the plasmonic peak larger than Au NPs, again in agreement with experiments.\cite{chen2008shape,ghosh2004solvent, zhang2016noble, lee2011refractive} Such shift is due to the refractive index sensitivity (RIS) of the LSP, a quantity that is commonly exploited in plasmonic colorimetric sensors.\cite{malinsky2001chain,mock2003local,nath2004label,underwood1994effect} RIS is generally expressed as $\Delta \lambda$ (in nm) over refractive index unit - $\Delta \lambda$(nm)/RIU.

The computed \wfqfmu/FQ RIS values are 15.0 nm/RIU for Ag %(65.9 nm/RIU with the computed refractive index of water -- $n_{WAT}^{FQ}$ -- see Sec. \ref{sec:si_ref_indx} for more details) 
and 7.9 nm/RIU for Au, %(34 nm/RIU with computed $n_{WAT}^{FQ}$)
while the experimentally measured RIS are 120-160 nm/RIU (Ag) and 70 nm/RIU (Au).\cite{malinsky2001chain,mock2003local,nath2004label,underwood1994effect} Therefore, \wfqfmu/FQ absolute RIS values are systematically lower than experimental ones for Ag and Au NPs, however \wfqfmu/FQ can nicely reproduce the experimental Ag/Au ratio ($\sim$ 2).
%The computed \wfqfmu/FQ refractive index sensitivity (RIS - expressed in peak shift in nm over RI unit - $\Delta \lambda$nm/RIU) is 15.1 nm/RIU for Ag, and 7.9 nm/RIU for Au. Notably, these values are a factor ten lower than the reported values for Ag (120-160 nm/RIU) and Au (70 nm/RIU) spheres, but remarkably their ratio is matched ($\sim$ 2). QUESTO E' UN ' ALTERNATIVA
The discrepancy can be related to the FQ parameters employed for modeling water, which have not been specifically tuned to describe solvated NPs.\cite{rick1994dynamical} Nevertheless, although experimental absolute RIS values are underestimated by a factor of 10, the capability of \wfqfmu/FQ to match the Ag/Au sensitivity ratio is a remarkable feature of the model, which can be used to rationalize the optical behavior of plasmonic nanostructures with high RIS, with potential applications in sensor design. 

\subsection{Au@Ag core-shell colloidal NPs}
%\wfqfmu as a tool to engineer higher sensitivity LSP colorimetric sensors: the effect of chemical substitution on high RIS Au@Ag core-shell NPs.

By taking advantage of the favorable computational scaling of \wfqfmu/FQ, its atomistic nature, and its capability of correctly reproducing Ag/Au RI sensitivity ratio, the model is challenged to optimize the sensitivity of Au@Ag core-shell spherical NPs. In fact, colorimetric sensors exploiting the PRF shift of noble metals upon change of the local refractive index (RI) are widely used in biosensing \cite{lee2013highly,kim2018heteroassembled,loiseau2019core,piliarik2012high,mock2003local,underwood1994effect,steinbruck2011sensoric,rycenga2011controlling,haes2004unified,anker2008biosensing,stewart2008nanostructured}. Among the various substrates used for this technology, gold-silver core-shell (Au@Ag) NPs are the most employed \cite{ mao2017novel,dong2013plasmonic,sun2014use,hao2014high,guo2016plasmonic,fu2012effect} due to their high RIS.\cite{steinbruck2011sensoric,fu2012effect} By studying such structures, we aim to showcase the potentialities of \wfqfmu/FQ, open up to colorimetric LSP sensor design.

\begin{figure}[!h]
    \centering
    \includegraphics[width=0.5\textwidth]{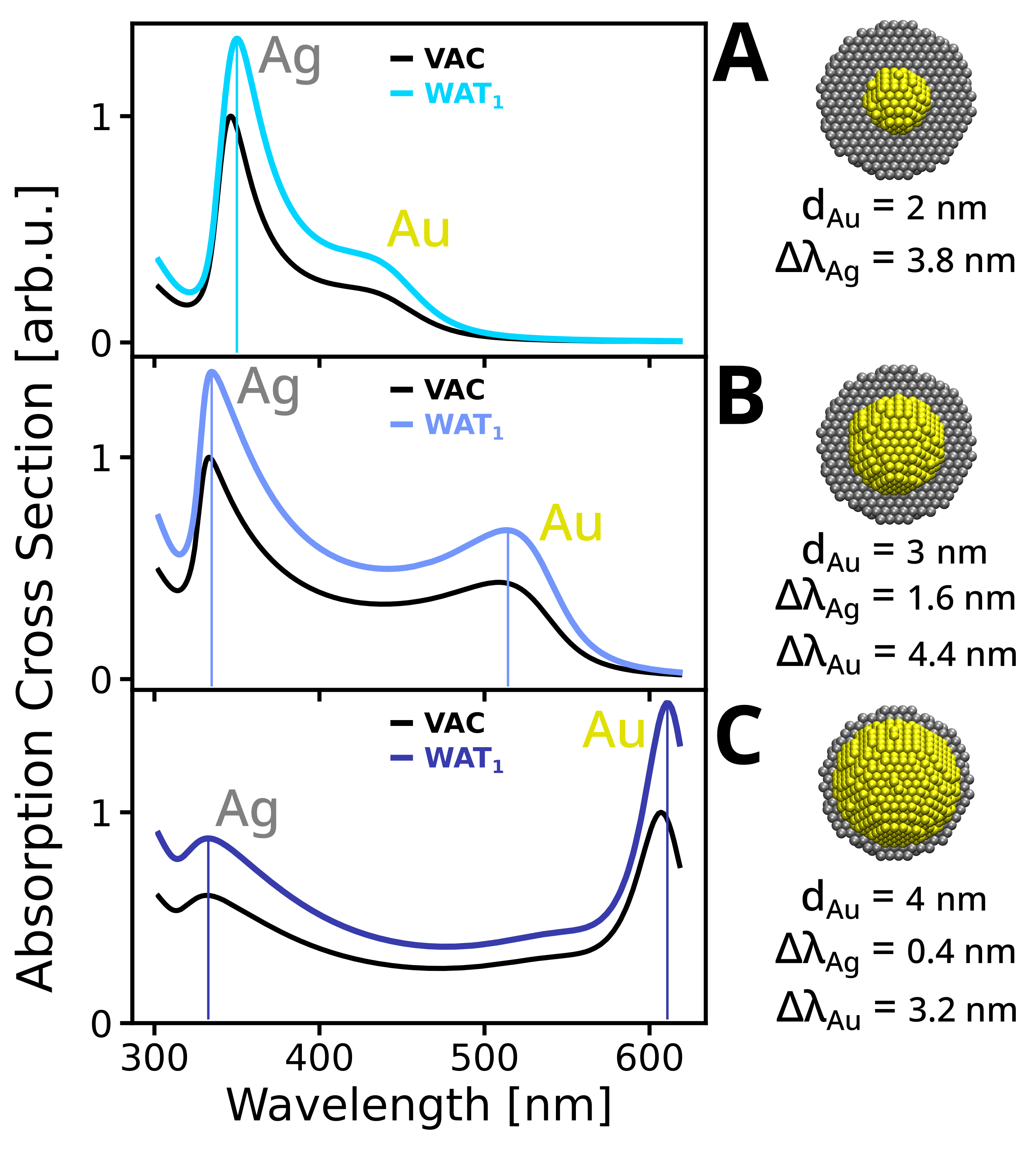}
    \caption{\wfqfmu/FQ absorption spectra of core-shell Au@Ag spherical NPs in vacuo (VAC) and in aqueous solution (WAT$_1$) as a function of the diameter of the Au core ($d_{Au}$: (A) 2.0 nm, (B) 3.0 nm and (C) 4.0 nm). The solvatochromic shifts of ``Ag'' ($\Delta \lambda_{Ag}$) and ``Au'' ($\Delta \lambda_{Au}$) peaks are reported in nm. All spectra are normalized to the maximum in vacuo.} %on the right of each panel together with the geometry of a single snapshot of each NP in vacuo (represented by slicing in half the Ag-shell). }
    \label{fig:core-shell-spheres}
\end{figure}

We first consider spherical Au@Ag core-shell NPs (diameter = 5 nm, 3851 atoms) characterized by an Au core of increasing size ($d_{Au}$=2, 3, and 4 nm -- see Fig. \ref{fig:core-shell-spheres}A,B,C), in aqueous solution. To sample the NP-solvent phase space, we perform a classical MD simulation of a single Ag NP in the aqueous solution. We then construct the Ag@Au core-shell bimetallic NP by properly substituting the metal atoms after the MD snapshot's extraction, which is a physically consistent procedure due to the almost equal lattice constants of the two metals.\cite{haynes2014handbook} On average, 6464 water molecules are considered in the FQ region and modeled employing the WAT$_1$ parameterization.\cite{rick1994dynamical}

In Fig. \ref{fig:core-shell-spheres}, \wfqfmu/FQ absorption spectra of Au@Ag core-shell NPs in vacuo (black) and water (blue) are reported. For all structures, absorption spectra in vacuo and solution feature two main bands at short ($\sim$ 350 nm) and long wavelengths (> 400 nm), in agreement with previous theoretical and experimental observations.\cite{steinbruck2011sensoric,szanto2021numerical,chen2012study,pena2011ag} Such peaks are located in the spectral region of the plasmon bands of pure Ag and Au spheres (see Fig. \ref{fig:spheres}A,B). For this reason, we call them ``Ag'' and ``Au'' peaks, respectively. Their position and intensity strongly depend on the diameter of the Au core both in the gas phase and solution. In particular, by first focusing on vacuo results, for Au-core thickness of 1 nm (Fig. \ref{fig:core-shell-spheres}A), the ``Ag'' peak, located around 346 nm, dominates the computed spectrum, while the ``Au'' band appears as a shoulder. By increasing the Au-core size, the ``Ag'' peak slightly blueshifts of about 10 nm while decreasing in intensity. On the contrary, the ``Au'' peak largely redshifts of about 150 nm, and significantly increases in intensity, becoming dominant for Au-core diameters of 4 nm (1 nm Ag-shell thickness). Including the solvent redshifts both ``Ag'' and ``Au'' peaks and increases their intensity. The shift differs for each peak and each Au-core diameter. In particular, all ``Ag'' peaks show lower solvatochromic shifts as compared to pure Ag spheres ($\Delta \lambda =$ 5.0 nm see Fig. \ref{fig:spheres}A), and increasing the Au-core size reduces the redshift of the ``Ag'' peak from $\Delta \lambda_{Ag} =$ 3.8 nm ($d_{Au}$ = 2 nm) to $\Delta \lambda_{Ag} =$ 0.4 nm ($d_{Au}$ = 4 nm). The shift of the ''Au'' peak is generally larger (e.g. $\Delta \lambda_{Au} =$ 4.4 nm for $d_{Au}$ = 2 nm) than that of a pure Au sphere ($\Delta \lambda =$ 2.6 nm, see Fig. \ref{fig:spheres}B). This suggests that for small NPs ($d \sim$ 5 nm), the ``Au'' peak RIS can be increased by coating the NP with an Ag layer. Remarkably, our findings align with previous theoretical and experimental observations.\cite{steinbruck2011sensoric,sun2014use,fu2012effect}. 

To showcase more potentialities of \wfqfmu/FQ, we now move to study the effect of alloying the Ag shell on the refractive index sensitivity, for which an atomistic picture is essential.\cite{nicoli2023fully} Note that, to the best of our knowledge, the use of atomistic modeling in this field has received only a little attention.\cite{szanto2021numerical,ma2015study,steinbruck2011sensoric} %To fill this gap, we leverage the atomistic nature of \wfqfmu/FQ to assess the effect of chemical substitution on Au@Ag core-shell NPs, showing how this could potentially enhance their RIS. 

\begin{figure}[!t]
    \centering
    \includegraphics[width = 0.5\textwidth]{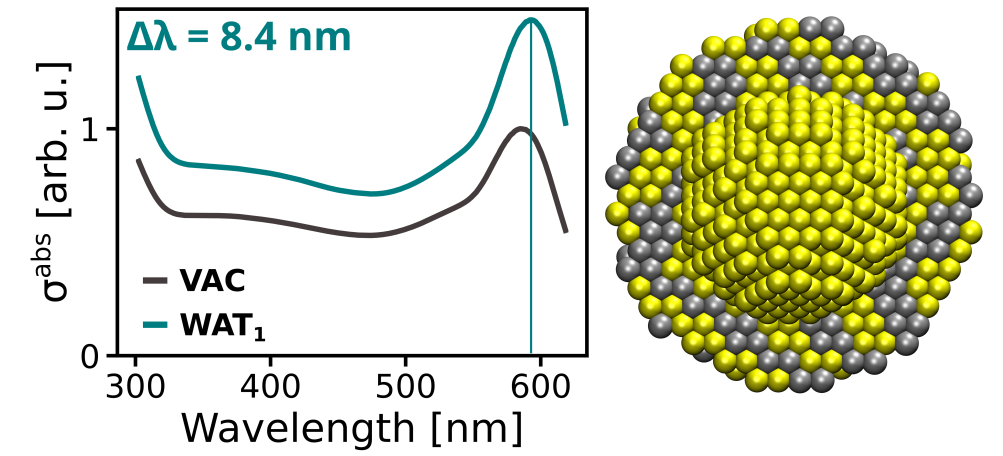}
    \caption{\wfqfmu/FQ absorption cross section ($\sigma^{abs}$) of core-shell Au@Ag spherical NP with an Au core diameter of 3 nm and featuring an alloyed (50\% Au/Ag) external layer of 1 nm (see right panel) in vacuo (VAC) and aqueous solution (WAT$_1$). $\Delta \lambda$ indicates the solvatochromic shift (in nm). All spectra are normalized to the maximum in vacuo.} %In the right panel, we report a graphical depiction of one of the 25 snapshots employed for the calculation of the spectrum in solution obtained by cutting in half the 50\% Au/Ag 1 nm thick external layer.}
    \label{fig:alloy}
\end{figure}

We consider a substantially more complex system composed of a spherical Au@Ag core-shell bimetallic NP (see Fig. \ref{fig:alloy}) of 5 nm of diameter (3851 atoms) with an Au core of diameter ($d_{Au}$) of 3 nm, and featuring a 50\% Ag 50\%Au random bimetallic layer of 1 nm. Such NP is solvated in aqueous solution, exploiting the same procedure discussed above.  
%To this end, we considered the spherical Au@Ag core-shell NP with higher RIS i.e. Au core of diameter ($d_{Au}$) of 3 nm (Fig. \ref{fig:core-shell-spheres}-B right panel), and we substituted randomically 50 \% of the silver shell atoms with gold, effectively realizing a 1 nm thick 50\% Ag 50\% Au random bimetallic layer. 
Fig. \ref{fig:alloy}, shows computed \wfqfmu/FQ absorption cross-section ($\sigma^{\mathrm{abs}}$) both in vacuo and water. The computed spectrum substantially deviates from the corresponding perfect core-shell system (Fig. \ref{fig:core-shell-spheres}B). In fact, the ``Au'' peak is shifted at higher wavelengths ($\sim$ 600 nm) and dominates the spectrum, while the ''Ag'' peak disappears in a dim spectral feature between 350 nm and  500 nm. Remarkably, the computed PRF shift of the gold peak ($\Delta \lambda_{Au}$) is 8.4 nm, almost double the original Au@Ag core-shell structure ($\Delta \lambda_{Au} = $ 4.4 nm, see Fig. \ref{fig:core-shell-spheres}B), and more than three times than a pure Au sphere of the same dimensions ($\Delta \lambda = $ 2.6 nm, see Fig. \ref{fig:spheres}). Remarkably, the same trend is also obtained for the computed \wfqfmu/FQ RIS sensitivity, which reaches 25.5 nm/RIU.
%(111 nm/RIU using $n_{WAT}^{FQ}$) . 
The obtained results thus suggest that the alloying of the Ag shell of an Au@Ag core shell can potentially improve the refractive index sensitivity of the system by almost a factor of 2. %Thus, the predictive capability of the enhancement of the RIS of a generic plasmonic NP makes \wfqfmu/FQ a cost-effective tool for the in-silico design of high-sensitivity LSP colorimetric sensors.

To finally showcase the flexibility of \wfqfmu/FQ, we solvate the NP in a 1:1 water-ethanol mixture (WAT-ETH). The even more complex chemical nature of this system enforces the need for a fully atomistic model to simulate its optical properties. To the best of our knowledge, \wfqfmu/FQ is the first fully atomistic model capable of calculating the response of a generic multimetallic plasmonic NP embedded in a multicomponent solvent.

\begin{figure}
    \centering
    \includegraphics[width = 0.5\textwidth]{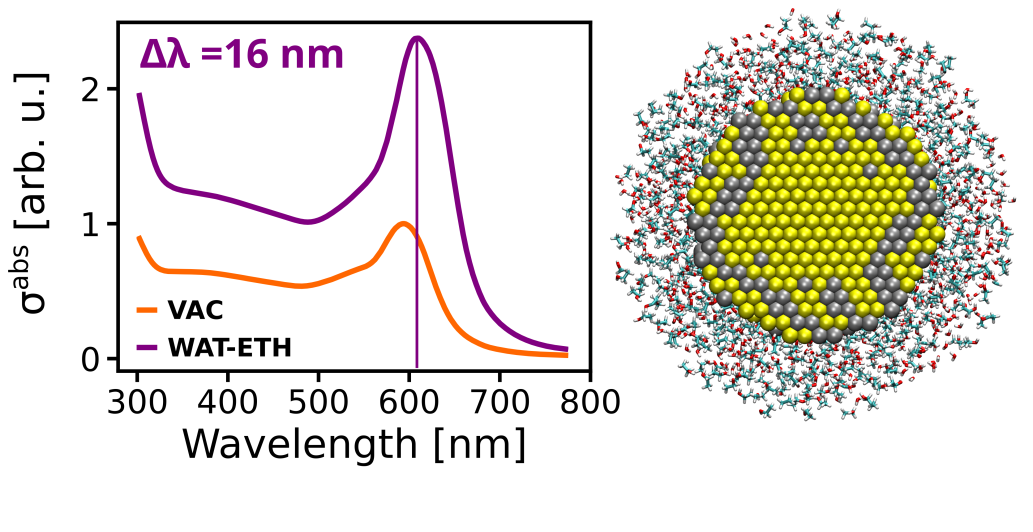}
    \caption{\wfqfmu/FQ absorption cross section ($\sigma^{abs}$) of core-shell Au@Ag spherical NP with an Au core diameter of 3 nm and featuring an alloyed (50\% Au/Ag) external layer of 1 nm (see right panel) in vacuo (VAC) and a 1:1 water-ethanol mixture (WAT-ETH). The vacuo-to-mixture solvatochromic shift is given in nm. All spectra are normalized to the maximum in vacuo.} %In the right panel, we report a graphical depiction of one of the structures employed for these calculations, obtained by slicing in half the structure for better visualization.}
    \label{fig:immagine-mix}
\end{figure}

To sample the NP-solvent phase space, we perform a classical MD simulation of a single Ag NP in the WAT-ETH mixture. We then construct the Ag@Au core-shell bimetallic NP by properly substituting the metal atoms after the MD snapshot extraction. The FQ parameterization employed for the water molecules is WAT$_1$, while for ethanol we exploit the parameters proposed in Ref. \citenum{ambrosetti2021quantum} (see Tab.S1 in the \sm). On average 1800 ETH and 1218 WAT molecules are included in the spherically-shaped snapshot (i.e. 19854 atoms in total). 

In Fig. \ref{fig:immagine-mix}, the absorption spectrum of the complex bimetallic Au/Ag NP both in vacuo (VAC) and solvated in the 1:1 water-ethanol mixture (WAT-ETH) is graphically depicted, together with the corresponding computed PRF shift ($\Delta \lambda$ in nm). Notably, absorption spectra both in vacuo and in solution present a sharp peak around 600 nm which can be assigned to the Au plasmonic dipolar band, and a shoulder at about 350-400 nm associated with the Ag plasmonic absorption. Solvent effects provided by the WAT-ETH mixture lead to a huge enhancement of the peak absorption (almost twice that in vacuo), which is also red-shifted by 16 nm. Considering a refractive index for the WAT-ETH mixture of 1.36,\cite{oelke1936refractive} a computed RIS of about 44.0 nm/RIU is obtained. Such a value is almost twice that obtained in pure water, highlighting a non-trivial RIS dependence on the solvent composition.  

% However, thanks to the systematic nature of such underestimation of the RI-sensitivities (same factor for all peaks), \wfqfmu/FQ is still capable of providing a platform for the estimation of solvent-induced PRF shifts and RI-sensitivity optimization, even for complex atomistic geometries.

\section{Summary and Conclusions}

We have presented a novel fully atomistic multiscale classical model, named \wfqfmu/FQ, which is capable of simulating the optical properties of real-size plasmonic colloidal nanoparticles (NPs) of a generic chemical nature. \wfqfmu/FQ is a multiscale model based on the mutual electrostatic interaction between the solvent and the NP. More specifically, the interaction of the plasmonic substrate with the external optical field is modeled employing \wfqfmu,\cite{giovannini2022we,nicoli2023fully} where each atom is endowed with a frequency-dependent charge and dipole, modeling intraband and interband plasmon decaying mechanisms, respectively. The solvent environment is considered transparent to the optical radiation and its instantaneous polarization is modeled through the polarizable FQ force field.\cite{rick1994dynamical,rick1995fluctuating,rick1996dynamical}

\wfqfmu/FQ has been challenged to reproduce reference TD-DFTB/FQ values, showing an almost perfect match with the TD-DFTB/FQ vacuo-to-water plasmon resonance frequency (PRF) shift of a small silver cluster (\agcluster), with a substantial 99.4 \% speed-up of the calculation. %Such results are even more remarkable since the dimensions of the studied Ag NP are within the limit of quantum confinement. 
Then, we have showcased the capabilities of \wfqfmu/FQ by simulating the optical properties of real-size homogeneous Ag and Au spherical NPs ($\sim$ 5 nm of diameter, 3851 metal atoms with 6464 water molecules), highlighting how the computed ratio between the refractive index sensitivities of Au and Ag NPs matches the experiments. 
Remarkably, \wfqfmu/FQ can also be used to study the sensitivity of colorimetric LSP sensors, as demonstrated by the chemical substitution of Au atoms in Au@Ag core-shell NPs, which can potentially enhance the sensitivity by a factor 2 $\sim$ 3. Finally, the flexibility of \wfqfmu/FQ is validated by simulating the absorption spectrum of a bimetallic Ag/Au NP solvated in a 1:1 water-ethanol mixture. Remarkably, the model can be applied to any solvent or solvent mixtures, including green solvents,\cite{atilhan2018molecular} pending a reliable parametrization of the FQ force field.\cite{ambrosetti2021quantum}

In conclusion, \wfqfmu/FQ is the first fully atomistic classical model capable of providing a platform for the calculation of LSP shifts of plasmonic NPs with the accuracy of \textit{ab-initio} methodologies for systems in the quantum confinement size region, but with a computational cost that consents its application to realistic-sized colloidal NPs. Such a development can potentially pave the way for future in-silico rational design of colorimetric sensors. 

\section*{Acknowledgments}

We gratefully acknowledge the Center for High-Performance Computing (CHPC) at SNS for providing the computational infrastructure.

%\subsection*{Funding Sources}

%This work has received funding from the European Research Council (ERC) under the European Union’s Horizon 2020 research and innovation programme (grant agreement No. 818064). SC thanks the Horizon 2020 EU grant ProID (grant agreement No. 964363) for funding.

\section*{Supporting Information}

Details on \wfqfmu/FQ, computational details, convergence of \wfqfmu/FQ spectra as a function of the radius of the solvent droplet and number of snapshots, MD analysis for spherical NPs (diameter = 5 nm) in solution.

\medskip

\bibliography{biblio}
\end{document}